\documentclass{entcs} 

\usepackage{entcsmacro}
\usepackage{times}
\usepackage[english]{babel}
\usepackage{graphics}
\usepackage{url}
\usepackage{lhs2TeXPreamble}
\usepackage{boxedminipage}

\def\lastname{L{\"a}mmel}

\newcommand{\cod}{\kappa}

\renewcommand{\qquad}{\hspace{55\in}}

\newcommand{\snipheader}[1]{#1\vspace{-15\in}}
\newcommand{\mysection}[1]{\medskip

\noindent
{\large\textbf{#1}}}

\newcommand{\myparagraph}[1]{\medskip

\noindent\textbf{#1}\ }

\newcommand{\myfigure}[4]{
\begin{figure}[#4]
{\small
\framebox{\parbox{.962\textwidth}{#1\hfill\vspace{-80\in}}}%
\vspace{-10\in}\caption{#2}
\label{#3}}
\end{figure}
}

\begin{document}

\begin{frontmatter}

\title{The Sketch of a Polymorphic Symphony}

\vspace{-42\in}

\author{Ralf L{\"a}mmel}
\address{Email:
         \href{mailto:ralf@cs.vu.nl}%
         \texttt{\normalshape ralf@cs.vu.nl}\\
         CWI, Kruislaan 413, NL-1098 SJ Amsterdam\\
         Vrije Universiteit, De Boelelaan 1081a, NL-1081 HV Amsterdam}

\begin{abstract}
In previous work, we have introduced \emph{functional} strategies,
that is, first-class generic functions that can traverse into terms of
any type while mixing uniform and type-specific behaviour. In the
present paper, we give a detailed description of one particular
Haskell-based model of functional strategies. This model is
characterised as follows. Firstly, we employ \emph{first-class
polymorphism} as a form of second-order polymorphism as for the mere
types of functional strategies. Secondly, we use an encoding scheme of
run-time \emph{type case} for mixing uniform and type-specific
behaviour. Thirdly, we base all traversal on a fundamental combinator
for \emph{folding over constructor applications}.

Using this model, we capture common strategic traversal schemes in a
highly parameterised style. We study two original forms of
parameterisation. Firstly, we design parameters for the specific
control-flow, data-flow and traversal characteristics of more concrete
traversal schemes. Secondly, we use overloading to postpone commitment
to a specific type scheme of traversal. The resulting portfolio of
traversal schemes can be regarded as a challenging benchmark for
setups for typed generic programming.

The way we develop the model and the suite of traversal schemes, it
becomes clear that parameterised + typed strategic programming is best
viewed as a potent combination of certain bits of parametric,
intensional, polytypic, and ad-hoc polymorphism.
\end{abstract}

\end{frontmatter}

%%%%%%%%%%%%%%%%%%%%%%%%%%%%%%%%%%%%%%%%%%%%%%%%%%%%%%%%%%%%%%%%%%%%%%%%%%%%%

\mysection{Arrangement}

\begin{figure}[ht!]
\begin{boxedminipage}{\hsize}
\parbox{.30\textwidth}{
\resizebox{!}{450\in}%
{\includegraphics{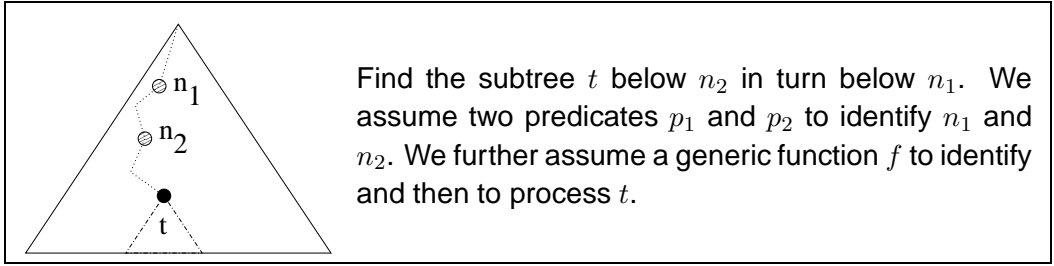}}}\hfill\parbox{.67\textwidth}{%
{\small \textsf{Find the subtree $t$ below $n_2$ in turn below $n_1$.
We assume two predicates $p_1$ and $p_2$ to identify $n_1$ and $n_2$.
We further assume a generic function $f$ to identify and then to
process $t$.}}}\vspace{-20\in}%
\end{boxedminipage}
\caption{Processing a subtree that is reachable via a path}
\label{F:tree}
\medskip
\end{figure}

\myparagraph{The running example}
Given is a tree, say a term. We want to operate on a certain
subterm. We search for the subterm $t$ in some traversal order, be it
top-down and left-to-right. We want to further constrain $t$ in the
sense that it should occur on a certain path, namely below $k$ nodes
the fitness of which is determined by predicates. Once we found $t$
below certain nodes $n_1$, \ldots, $n_k$, we either want to select $t$
as is, \emph{or} we want to compute some value from $t$, \emph{or} we
want to transform $t$ in its context in the complete tree. This
traversal scenario is illustrated in Fig.~\ref{F:tree} for $k=2$. This
sort of problem is very common in programming over tree and graph
structures, e.g., in adaptive object-oriented
programming~\cite{Lieberherr96}. One might consider additional forms
of conditions, e.g., certain kinds of nodes that should \emph{not} be
passed, or further forms of computation, e.g., \emph{cumulative}
computation along the path.

\myparagraph{Sample code}
In the present paper, we refrain from discussing real-world
application snippets but we sketch an illustrative instance of the
running example in Fig.~\ref{F:running}. As one can see, we assume a
combinator \emph{belowlist} that captures the described scheme of path
traversal. In the figure, we define a function \emph{test42} to
traverse into a sample term \emph{term1} in terms of
\emph{belowlist}. The traversal looks for a subterm of type
\emph{SortB} to extract its integer component if it occurs on a path
with two \emph{SortB} subterms with the integer components 1 and
3. There are a few blind spots in the figure (cf.\ ``\ldots'') which
we will resolve in the course of the concert. As the following Hugs
session shows, the traversal yields 42:

\myfigure{%
\noindent
\snipheader{A traversal scheme for the introductory problem}\begin{tabbing}
\qquad\=\hspace{\lwidth}\=\hspace{\cwidth}\=\+\kill
$\Varid{f}\mathbin{`\Varid{belowlist}`}\Varid{ps}\mathrel{=}\mathbin{...}$
\end{tabbing}\snipheader{A sample system of two datatypes}\begin{tabbing}
\qquad\=\hspace{\lwidth}\=\hspace{\cwidth}\=\+\kill
$\mathbf{data}\;\Conid{SortA}\mathrel{=}\Conid{SortA1}\;\Conid{SortB}\mid \Conid{SortA2}$\\
$\mathbf{data}\;\Conid{SortB}\mathrel{=}\Conid{SortB}\;\Conid{Int}\;\Conid{SortA}$
\end{tabbing}\snipheader{A test term for traversal}\begin{tabbing}
\qquad\=\hspace{\lwidth}\=\hspace{\cwidth}\=\+\kill
$\Varid{term1}\mathrel{=}\Conid{SortA1}\;(\Conid{SortB}\;\mathrm{0}\;(\Conid{SortA1}\;(\Conid{SortB}\;\mathrm{1}\;(\Conid{SortA1}\;(\Conid{SortB}\;\mathrm{2}\;($\\
$\phantom{\Varid{term1}\mathrel{=}\mbox{}}\Conid{SortA1}\;(\Conid{SortB}\;\mathrm{3}\;(\Conid{SortA1}\;(\Conid{SortB}\;\mathrm{42}\;\Conid{SortA2})))))))))$
\end{tabbing}\snipheader{Extract the integer from a \emph{SortB} term; fail for other sorts}\begin{tabbing}
\qquad\=\hspace{\lwidth}\=\hspace{\cwidth}\=\+\kill
$\Varid{sortb2int}\mathrel{=}\mathbin{...}$
\end{tabbing}\snipheader{Insist on a \emph{SortB} term with a specific integer}\begin{tabbing}
\qquad\=\hspace{\lwidth}\=\hspace{\cwidth}\=\+\kill
$\Varid{sortbEqInt}\;\Varid{i}\mathrel{=}\mathbin{...}$
\end{tabbing}\snipheader{An actual traversal}\begin{tabbing}
\qquad\=\hspace{\lwidth}\=\hspace{\cwidth}\=\+\kill
$\Varid{test42}\mathbin{::}\Conid{Maybe}\;\Conid{Int}$\\
$\Varid{test42}\mathrel{=}(\Varid{f}\mathbin{`\Varid{belowlist}`}[\mskip1.5mu \Varid{p1},\Varid{p2}\mskip1.5mu])\;\Varid{term1}$\\
$\hskip0.50em\relax\mathbf{where}\;\Varid{f}\mathrel{=}\Varid{sortb2int}$\\
$\hskip0.50em\relax\phantom{\mathbf{where}\;\mbox{}}\Varid{p1}\mathrel{=}\Varid{sortbEqInt}\;\mathrm{1}$\\
$\hskip0.50em\relax\phantom{\mathbf{where}\;\mbox{}}\Varid{p2}\mathrel{=}\Varid{sortbEqInt}\;\mathrm{3}$
\end{tabbing}%
}{%
Illustrative Haskell code for the running example%
}{%
F:running}{b!}

\begin{quote}
\mbox{}\vspace{-40\in}\\
\ \texttt{Main$>$ test42}\\%
\ \texttt{Just 42}\\%
\ \texttt{Main$>$}
\end{quote}

%%%%%%%%%%%%%%%%%%%%%%%%%%%%%%%%%%%%%%%%%%%%%%%%%%%%%%%%%%%%%%%%%%%%%%%%%%%%%%%

\myparagraph{Side conditions}
The above traversal scenario should be complemented by a few useful
side conditions. We assume that (i) we deal with \emph{many sorts},
say user-supplied systems of named, mutually recursive datatypes such
as syntaxes and formats.  So a traversal will encounter terms of
different types. We further require that (ii) traversal schemes are
\emph{statically type-safe}, that is, a traversal always delivers a
type-correct result (if any) without even attempting the construction
of any ill-typed term. Furthermore, we insist on (iii) a
\emph{parameterised} solution where the overall traversal scheme is
strictly separated from problem-specific ingredients. Also, the kind
of processor (recall selection vs.\ computation vs.\ transformation)
should not be anticipated. Recall that the actual traversal in
Fig.~\ref{F:running} is indeed synthesised by passing predicates and a
processor to a presumably overloaded traversal scheme
\emph{belowlist}. Last but not least, we require that (iv) a proper
combinator approach is adopted where one can easily compose traversals
and schemes thereof. As for the assumed \emph{belowlist} combinator,
its definition should be a concise and suggestive one-liner based on
more fundamental traversal combinators. With combinators, we can
easily toggle variation points of traversal, e.g., top-down vs.\
bottom-up, left-to-right vs.\ right-to-left, first match vs.\ all
matches, and so on.

%%%%%%%%%%%%%%%%%%%%%%%%%%%%%%%%%%%%%%%%%%%%%%%%%%%%%%%%%%%%%%%%%%%%%%%%%%%%%%%

\myparagraph{Conducted by Strafunski}
If \emph{any} of the above side conditions is given up, the treatment
of the scenario becomes less satisfying. If you do not insist on
static type safety, then you can resort to Prolog, or preferably to
Stratego~\cite{VBT98} which supports at least limited type
checks. Types are however desirable to discipline the instantiation of
traversal schemes. If you even want to do away with parameterisation
and combinator style, then XSLT is your language of choice. If types
appeal to you, then you might feel tempted to consider polytypic
functional programming~\cite{JJ96}.  However, corresponding language
designs do not provide support for generic programming in a combinator
style because polytypic values are not first-class citizens. If you
are as demanding and simplistic as we are, then you use the
\emph{Strafunski}-style of generic programming in Haskell as
introduced by the present author and Joost Visser
in~\cite{LV02-PADL}.\footnote{URL \url{http://www.cs.vu.nl/Strafunski}
where \url{Stra} refers to strategies as in strategic term
rewriting~\cite{BKKR01,VBT98}, \url{fun} refers to functional
programming, and their harmonious composition is a homage to the music
of Igor Stravinsky.}  The style is centred around the notion of
functional strategies~---~\emph{first-class} generic functions that
can \emph{traverse} into terms of any type while mixing uniform and
\emph{type-specific} behaviour.

%%%%%%%%%%%%%%%%%%%%%%%%%%%%%%%%%%%%%%%%%%%%%%%%%%%%%%%%%%%%%%%%%%%%%%%%%%%%%%%

\myparagraph{The movements}
No previous knowledge of strategic programming is required to enjoy
this symphony. We develop a model of functional strategies based on
Haskell~98~\cite{HaskellReport} extended with first-class
polymorphism~\cite{Jones97}. The symphony consists of four
movements. In Sec.~1, the type \emph{Strategic} of functional
strategies is initiated and inhabited with parametrically
polymorphic~\cite{Reynolds83,Wadler89} combinators.  In Sec.~2, we
integrate a combinator \emph{adhoc} to update functional strategies
for a specific type. This combinator relies on run-time type
case~---~a notion that was studied in the context of intensional
polymorphism~\cite{HM95,Weirich02}. In Sec.~3, we integrate a
fundamental combinator \emph{hfoldr} to perform primitive portions of
traversal by folding over the immediate subterms of a term without
anticipation of recursion.  This kind of folding can be regarded as a
variation on polytypism~\cite{JJ96}. In Sec.~4, we use overloading,
say ad-hoc polymorphism~\cite{WB89,Jones95}, to treat different types
of traversal (recall selection vs.\ computation vs.\ transformation)
in a uniform manner. At this level of genericity, we succeed in
defining a portfolio of highly parameterised traversal schemes,
including the scheme \emph{belowlist} that is needed in the running
example. This portfolio demonstrates the expressiveness and
conciseness of functional strategies.%
\footnote{%
The code shown in the paper is fully operational in Haskell (tested
with Hugs~98; Dec.\ 2001 version). A paper-specific \emph{Strafunski}
distribution with illustrative examples can be downloaded from the
paper's web-site \url{http://www.cwi.nl/~ralf/polymorphic-symphony/}.
This distribution also comes with a strategic Haskell program that
provides generative tool support for strategic programming against
user-supplied datatypes.}

%%%%%%%%%%%%%%%%%%%%%%%%%%%%%%%%%%%%%%%%%%%%%%%%%%%%%%%%%%%%%%%%%%%%%%%%%%%%%%%

\section{Moderato}

In this section, we initiate the type of generic functions that model
functional strategies. For the sake of a systematic development, we
will first focus on the most simple, i.e., the \emph{parametrically
polymorphic}~\cite{Reynolds83,Wadler89} dimension of the corresponding
type scheme. In the subsequent two movements, we go beyond this
boundary by adding expressiveness for run-time type case and generic
term traversal.

\myfigure{%
\input{snip/generic.math}%
}{%
Functional strategies as first-class polymorphic functions%
}{%
F:generic}{}

\myparagraph{First-class polymorphic functions}
In Fig.~\ref{F:generic}, we use a type synonym \emph{Parametric} with
two parameters $\alpha$ and $\cod$ to capture the general type scheme
$\alpha \to \cod\ \alpha$ of functional strategies. As one can see,
the parameter $\alpha$ corresponds to the domain of the function type,
and $\cod$ is used to construct the co-domain from $\alpha$. Common
options for the co-domain type constructor $\cod$ are the identity
type constructor $I$ and the constant type constructor $C$ as defined
in the figure, too. Based on the type scheme \emph{Parametric}, we
define the ultimate datatype \emph{Strategic} that captures the
polymorphic function type for functional strategies.  Polymorphism is
expressed via the universal quantifier ``$\forall$'', and the fact
that we ultimately need to go beyond parametric polymorphism is
anticipated via a class constraint ``$\emph{Term}\ \alpha$'' for
`strategic polymorphism'. Note that the ``$\forall$'' occurs in the
scope of the \emph{component} of the constructor $G$. This form of
second-order polymorphism, where polymorphic entities are wrapped by
datatype constructors, is called first-class
polymorphism~\cite{Jones97}, and it is a common language extension of
Haskell~98. (Implicit) top-level universal quantification would be
eventually insufficient to construct strategy combinators.

\myparagraph{Strategic type schemes}
In Fig~\ref{F:generic}. we also define the identity type constructor
$I$, the constant type constructor $C$, and the constructor $S$ for
sequential composition of type constructors. Based on these, we derive
three classes \emph{MG}, \emph{TP}, and \emph{TU} of generic functions
from \emph{Strategic}. The synonym \emph{MG} captures monadic
strategies.  Monadic style~\cite{Wadler92} is favoured here because
that way we can deal with effects during traversal, e.g., success and
failure, state, or nondeterminism. We could also consider extra
variants for non-monadic strategies but we instead assume that the
trivial identity monad is used for that purpose. The synonym \emph{TP}
instantiates \emph{MG} via $I$ so that \emph{T}ype-\emph{P}reserving
strategies are identified. The synonym \emph{TU} instantiates
\emph{MG} via $C$ so that \emph{T}ype-\emph{U}nifying strategies are
identified, that is, polymorphic functions with a fixed result
type. At the bottom of Fig.~\ref{F:generic}, we provide trivial
definitions of `application' combinators. The combinator \emph{apply}
complements ordinary function application by some unwrapping that
accounts for first-class polymorphism (cf.\ \emph{unG}), and for
type-constructor composition (cf.\ \emph{unS}). The combinators
\emph{applyTP} and \emph{applyTU} specialise \emph{apply} for
\emph{TP} and \emph{TU} to hide the employment of the datatypes $I$
and $C$ for co-domain construction. Their result types point out the
basic type schemes for \emph{TP} and \emph{TU} without any noise.%
\footnote{%
We should revise our sample code from Fig.~\ref{F:running} to use
\emph{applyTU} in the synthesis of the traversal:
\input{snip/revise.math}}

\myfigure{
\input{snip/constants.math}
}{Parametrically polymorphic strategies}{F:constants}{}

\myparagraph{Nullary combinators}
Let us start to inhabit the strategy types while restricting ourselves
to parametrically polymorphic inhabitants. In Sec.~2 and Sec.~3, we
will inhabit the type \emph{Strategic} in other ways by employing
designated combinators for `strategic polymorphism'. If we neglect the
monadic facet of functional strategies for a second, then there are
few parametrically polymorphic inhabitants: the identity function for
the type-preserving scheme, and constant functions for the
type-unifying scheme. This is generalised for the monadic setup in
Fig.~\ref{F:constants}. For clarity, we define the function
\emph{para} that approves a parametrically polymorphic function as a
member of \emph{Strategic}. While its definition coincides with the
constructor $G$, its type insists on a parametrically polymorphic
argument (no mentioning of the \emph{Term} class). Using \emph{para},
we can define the identity strategy \emph{idTP} and a combinator
\emph{constTU} for constant strategies. Relying on the extended monad
class \emph{MonadPlus}, we can also define the always failing strategy
\emph{fail} in terms of the member \emph{mzero} denoting failure,
e.g., \emph{Nothing} in the case of the \emph{Maybe} instance of the
\emph{MonadPlus} class. The inhabitant \emph{fail} is meant to
illustrate that functional strategies might exhibit success and
failure behaviour, or they even might be non-deterministic depending
on the actual choice of the \emph{MonadPlus} instance.

\myfigure{
\input{snip/combinators.math}
}{Composition combinators for functional
strategies}{F:combinators}{}

\myparagraph{Binary combinators}
In Fig.~\ref{F:combinators}, we define three prime combinators to
compose functional strategies. The combinator \emph{seq} lifts monadic
sequencing of function applications to the strategy level (cf.\
``$\bind$''). The combinator \emph{pass} composes two strategies which
share a term argument and where the result of the first strategy is
passed as additional input to the second strategy. The combinator
\emph{choice} composes alternative function applications. The actual
kind of choice depends on the (extended) monad which is employed in
the definition (cf.\ class \emph{MonadPlus} and its member
\emph{mplus}), e.g., the \emph{Maybe} monad for partiality or the
\emph{List} monad for multiple results. In the definitions of the
binary combinators, we use \emph{apply} to deploy the polymorphic
function arguments, and we use $G$ to re-wrap the composition as a
polymorphic function of type \emph{Strategic}. In composing functional
strategies, we cannot use the disciplined \emph{para} as a substitute
for $G$ because, eventually, we want to compose functions that go
beyond parametric polymorphism.

\myparagraph{Rank-2 types}
A crucial observation is that the composition combinators \emph{seq},
\emph{pass} and \emph{choice} really enforce us to use some form of
second-order polymorphism~\cite{GirardPhD,Reynolds74}. Other function
combinators like ``$\circ$'' or ``$\bind$'' have rank-1 types, that
is, they are quantified at the top-level. This implies that they
(also) accept monomorphic function arguments. By contrast, the types
of \emph{seq}, \emph{pass} and \emph{choice} have rank-2 types, that
is, they are quantified \emph{argument-wise}. These types model
insistence on polymorphic function arguments as required for
\emph{generic} programming in a combinator style. Rank-2 types are not
just needed for the binary composition combinators, but also for the
upcoming traversal primitive, and for non-recursive and recursive
traversal combinators. Actually, generic traversal in a combinator
style necessitates rank-2 types. This is because, in general, a
traversal \emph{scheme} takes universally quantified arguments.
Nested quantification reflects that ingredients of traversals must be
applicable to subterms of any type.

%%%%%%%%%%%%%%%%%%%%%%%%%%%%%%%%%%%%%%%%%%%%%%%%%%%%%%%%%%%%%%%%%%%%%%%%%%%%%%%

\section{Larghetto}

At this point in our polymorphic concert, we still lack the original
expressiveness of functional strategies: the ability to perform
traversal into terms while mixing uniform and type-specific
behaviour. This section and the subsequent one are concerned with this
`strategic polymorphism'. In the present section, we provide a
combinator \emph{adhoc} for type-specific customisation of functional
strategies. In the next section, generic traversal into terms will be
enabled.

\myparagraph{Informal explanation}
The type of the \emph{adhoc} combinator is given at the top of
Fig.~\ref{F:adhoc}. The combinator allows us to update an existing
strategy for a specific type via a monomorphic function. We call this
idiom \emph{type-based function dispatch}. We also use the term
\emph{strategy update} because of the affinity to point-wise
modification of ordinary functions. In this sense, strategy update is
about ``type-wise'' modification of a polymorphic function. Here are
two samples of strategy update:

\input{snip/negbool.math}

\input{snip/recint.math}

\noindent
The strategy \emph{negate\_bool} behaves like \emph{idTP} most of the
time but it performs negation when faced with a Boolean (cf.\
``$\neg$'').  We use the \emph{Identity} monad. The strategy
\emph{return\_int} is meant to recognise integers, that is, it simply
fails if it does not find an integer. So we use the \emph{Maybe} monad
to deal with failure. Note that we use type-specialised variants
\emph{adhocTP} and \emph{adhocTU} as defined in
Fig.~\ref{F:adhoc}. They allow the strategic programmer to forget
about wrapping and unwrapping $I$, $C$, and $S$.

\medskip

\noindent
In Fig.~\ref{F:blind}, we use strategy update to resolve two blind
spots in Fig.~\ref{F:running}~---~the sample code for our running
example. The strategy \emph{sortb2int} extracts an integer from a term
of type \emph{SortB}. The strategy \emph{sortbEqInt} insists on a
specific extracted integer. This is a predicate-like strategy which
either returns $\mathit{Just}\,()$ or $\mathit{Nothing}$.

\myfigure{%
\input{snip/adhoc.math}}{%
Updating functional strategies by type-specific cases% 
}{%
F:adhoc%
}{t}

\myfigure{%
\input{snip/blind.math}}{%
Use of strategy update in the running example% 
}{%
F:blind%
}{t}

\myparagraph{Definition of strategy update}
Given a strategy \emph{poly} of type $\mathit{Strategic}\ \cod$, and a
function \emph{mono} of type $\mathit{Parametric}\ \alpha\ \cod$, the
updated function $\mathit{adhoc}\ \mathit{poly}\ \mathit{mono}$
branches according to the type $\alpha'$ of an eventual input value
$v$: if $\alpha=\alpha'$, then \emph{mono} is applied to $v$,
otherwise \emph{poly}. That is:
\[
\mathit{adhoc}\ \mathit{poly}\ \mathit{mono}\ v\ =\ \left\{%
\begin{array}{ll}
\mathit{mono}\ v, & \mbox{if type of $v$ = domain of
$\mathit{mono}$}\vspace{-30\in}\\
\mathit{poly}\ v, & \mbox{otherwise}
\end{array}
\right.
\]

\medskip

\noindent
Strategy update can be considered as a simple and very disciplined
form of type case at run-time. Such type case was studied in the
context of intensional polymorphism~\cite{HM95,Weirich02}. One might
feel tempted to compare strategy update to dynamic
typing~\cite{ACPP91,ACPR92} but note that first-class polymorphic
strategies, as designed in the present paper, operate on terms of
algebraic datatypes as opposed to `dynamics'. Dynamic typing can be
employed, however, in other models of functional strategies, e.g., in
the way described in~\cite{LV02-PADL}.

\myfigure{%
\input{snip/typecase.math}}{%
Per-datatype support for strategy update%
}{%
F:typecase%
}{b!}

\myparagraph{Encoding scheme}
A strategic programmer considers the \emph{adhoc} combinator as a
primitive. In Haskell, we cannot define the \emph{adhoc} combinator
once and for all, but it can be supported per datatype. This is
precisely what the \emph{Term} class is needed for in the definition
of \emph{Strategic} (recall Fig.~\ref{F:generic}). So we basically
place the \emph{adhoc} combinator in the \emph{Term} class with the
provision to add another member for generic traversal later. There are
several ways to encode strategy update. We will explain here a scheme
that is inspired by type-safe cast as of~\cite{Weirich00}. In fact, we
revise this scheme to model a \emph{single-branching type case} (as
opposed to a plain type cast) while assuming \emph{nominal type
analysis} (as opposed to a structural one).

\medskip

\noindent
The scheme is illustrated in Fig.~\ref{F:typecase}. In the \emph{Term}
class, we place a primed member $\mathit{adhoc}'$ with an implicitly
quantified type. This is necessary because we need to link the type
parameter $\alpha$ of the class declaration to the type of
$\mathit{adhoc}'$. In the type of $\mathit{adhoc}'$, the parameter
$\alpha$ acts as a place-holder for the term type of the updating
monomorphic function, and there is an extra parameter $\alpha'$ for
the term type of the input term. The first-class polymorphic
\emph{adhoc} is easily derived from $\mathit{adhoc}'$ via wrapping
with $G$ as shown in the figure. The overall idea for the
implementation of $\mathit{adhoc}'$ is to compare the term type
$\alpha$ of the update with the term type $\alpha'$ of the ultimate
input term. To this end, we need to encode a kind of decision matrix
to check for $\alpha=\alpha'$ for all possible combinations of term
types. The dimension for $\alpha$ is taken care of by overloading
$\mathit{adhoc}'$ in $\alpha$. The dimension for $\alpha'$ is taken
care of by extra helper members~---~one for each type. These helper
members model strategy update for a specific type of update while they
are overloaded in the input term type. This is exemplified in the
figure with the helpers \emph{int}, \emph{sorta}, and \emph{sortb} for
the sample term types in our running example (cf.\
Fig.~\ref{F:running}). It makes sense to initialise the definitions of
the helper members to resort to \emph{poly} by default. In any given
instance of the \emph{Term} class, we need to define
$\mathit{adhoc}'$, and we override the default definition for the
specific helper member that handles the instance's type. The
definition of $\mathit{adhoc}'$ simply dispatches to the helper
member. The helper member is redefined to dispatch to \emph{mono}
instead of \emph{poly}. This is again illustrated for the sample term
types in our running example. Note that we are faced with a
closed-world assumption because we need one member per term type in
our strategic program (see~\cite{LV02-PADL} for another model).

\myparagraph{Mixture of specificity and genericity}
The contribution of strategy update is the following. It allows us to
lift type-specific behaviour in a type-safe and transparent way to the
strategy level. Note that parametric polymorphism only works the other
way around: a polymorphic function can be applied to a value of a
specific type, and in fact, the actual behaviour will be entirely
uniform, independent of the specific
type~\cite{Reynolds83,Wadler89}. In generic programming with
traversals, type-specific customisation is indispensable because a
traversal is typically concerned with problem-specific datatypes and
constructors. At a more general level, one can say that the
\emph{adhoc} combinator implements the following admittedly very much
related requirements regarding the tension between genericity and
specificity:
\begin{description}
\item[(i)] Separability of generic and type-specific behaviour
\item[(ii)] Composability of generic and type-specific behaviour
\item[(iii)] First-class status of updatable generic functions
\end{description}
In functional programming, the common approach to the definition of
polymorphic functions with type-specific behaviour is to use ad-hoc
polymorphism, say type and constructor classes~\cite{WB89,Jones95}
as supported in Haskell. This approach does not meet the above
requirements because a class and its instances form a declaration
unit. Any new combination of generic and type-specific behaviour has
to be anticipated by a dedicated class. Hence, strategy update is
complementary.

%%%%%%%%%%%%%%%%%%%%%%%%%%%%%%%%%%%%%%%%%%%%%%%%%%%%%%%%%%%%%%%%%%%%%%%%%%%%%%%

\section{Allegretto}

Ultimately, functional strategies are meant to perform term traversal.
Hence, we need to add another bit of polymorphism. In fact, the kind
of term traversal that we envisage can be regarded as a variation on
polytypism~\cite{JJ96}. That is, recursion into terms is performed on
the basis of the structure of the underlying algebraic
datatypes. First we will briefly categorise traversal approaches in
previous work. Then we will generalise one of the approaches to
contribute a very flexible means of traversal to the ongoing
reconstruction of functional strategies.

\myparagraph{Traversal approaches}
We identify the following categories:
\begin{description}
\item[(i)] Recursive traversal schemes~\cite{MFP91,CS92,Thompson97,LVK00,BKV01}
\item[(ii)] One-layer traversal
combinators~\cite{VBT98,Laemmel02-TGT,LV02-PADL}
\item[(iii)] Traversal by type induction~\cite{JJ96}
\item[(iv)] Traversal at a representation-type level~\cite{BKKR01}
\end{description}
(i) Recursive traversal schemes, e.g., generalised folds or iterators
over data structures, follow a fixed scheme of recursion into terms,
and the style of composition of intermediate results is also
intertwined with the recursion scheme. (ii) One-layer traversal
combinators add flexibility because they do not recurse into terms by
themselves but they rather capture single steps of traversals. Hence,
different recursion schemes and different means to compose
intermediate results can still be established by plain (recursive)
function definition. The value of one-layer traversal combinators was
identified in the context of strategic term rewriting~\cite{VBT98} in
the application domain of program transformation. (iii) Polytypic
programming suggests that a traversal can also be viewed as a function
that is defined by induction on the argument type. So one can capture
term traversal schemes by polytypic definitions. However, these
definitions could not be used in a combinator style because existing
language designs do not cover `rank-2 polytypism', that is, one cannot
pass one polytypic function to another one. (iv) Finally, a programmer
can resort to a universal representation type to perform
traversal. Here we assume that the programmer is provided with
implosion and explosion functionality to mediate between
programmer-supplied term types and the representation type. While
unrestricted access to a representation type corresponds to a less
safe way of generic programming due to the potential of implosion
problems, \emph{implementations} of (i)--(iii) might very well be
based on a representation type. To give an example,
in~\cite{LV02-PADL}, traversal combinators are implemented using a
representation type without any exposure to the generic
programmer. Hence, implosion safety can be certified.

\myparagraph{Right-associative, heterogeneous fold}
In this symphony, we basically follow the one-layer
traversal-combinator approach, but we generalise it in the following
manner. Rather than making a somewhat arbitrary selection of traversal
combinators as in previous work~\cite{VBT98,Laemmel02-TGT,LV02-PADL},
we identify the fundamental principle underlying all one-layer
traversal. To this end, we define a primitive combinator for folding
over constructor applications. This kind of folding views the
immediate subterms of a term (say, its children) as a heterogeneous
list of terms.  Without loss of generality, we favour a
\emph{right-associative} fold in the sequel. We call the resulting
fold operation \emph{hfoldr} for right-associative, heterogeneous
fold. Let us recall how the ordinary \emph{foldr} folds over a
homogeneous list. Given the fold ingredients $f$ and $z$ for the
non-empty and the empty list form, the application of \emph{foldr} to
a list is defined as follows:
\[%
\mathit{foldr}\ f\ z\ [x_1,x_2,\ldots,x_n]\ =\ f\ x_1\ (f\ x_2\ (\cdots\ (f\ x_n\ z)\ \cdots))\]
The combinator \emph{hfoldr} folds over a constructor application
in nearly the same way:
\[%
\mathit{hfoldr}\ f\ z\ (C\ x_n\ \cdots\ x_2\ x_1)\ =\ %
f\ x_1\ (f\ x_2\ (\cdots\ (f\ x_n\ (z\ C))\
\cdots))\]
Note the order of the children in a constructor application. Due to
curried style, we have a kind of snoc-list, that is, the leftmost
child is the head of the list, and so on. Also note that the empty
constructor application $C$ is passed to $z$ so that the constructor
can contribute to the result of folding.

\myfigure{%
\input{snip/hfoldr.math}
}{%
Folding over constructor applications
}{%
F:hfoldr%
}{t}

\myfigure{%
\input{snip/hfoldrprime.math}
}{%
Per-datatype support for \emph{hfoldr}~---~sample instances%
}{%
F:hfoldr'%
}{t}

\myparagraph{Encoding scheme}
A strategic programmer considers \emph{hfoldr} as a primitive.  In
Haskell, we cannot define the \emph{hfoldr} combinator once and for
all, but it can be supported per datatype~---~as in the case of
strategy update. In Fig.~\ref{F:hfoldr}, we complete the \emph{Term}
class accordingly. We add a primed helper member $\mathit{hfoldr}'$ to
the class, and we define the actual combinator \emph{hfoldr} via
wrapping with $G$. Due to the employment of first-class polymorphism,
we need to pack the ingredients $f$ and $z$ for \emph{hfoldr} in a
datatype \emph{HFoldrAlg}. This is a deviation from the curried style
that is used for the rank-1 \emph{foldr} for lists. The datatype
constructor $A$ is used for packaging. The types of the two components
deserve some explanation. The first component, i.e., the one that
corresponds to $f$, is universally quantified in $\alpha$ and $\beta$
corresponding to the type of the heading child (i.e., the outermost
argument in the constructor application), and the type of the
constructor application at hand, respectively. The type of the
recursively processed tail is $\cod\ (\alpha \to \beta)$ because it
`lacks' a child of type $\alpha$.  Given a constructor application $C\
x_n\ \cdots\ x_2\ x_1$, the type variable $\alpha$ would be bound to
the respective types of the $x_i$ in the several folding steps while
the type $\beta$ variable would be bound to the corresponding
remainders of $C$'s type. The second component of $A$, i.e., the one
that corresponds to the base case $z$, is simply a parametrically
polymorphic function for processing the empty constructor
application. The per-datatype definition of \emph{hfoldr} is
illustrated in Fig.~\ref{F:hfoldr'} with the sample datatypes from our
running example. We basically need one equation for $\mathit{hfoldr}'$
of the above form for each specific datatype constructor.

\myfigure{%
\noindent
\snipheader{Process all children; preserve the outermost constructor}\begin{tabbing}
\qquad\=\hspace{\lwidth}\=\hspace{\cwidth}\=\+\kill
$\Varid{allTP}\mathbin{::}\Conid{Monad}\;\Varid{m}\Rightarrow \Conid{TP}\;\Varid{m}\to \Conid{TP}\;\Varid{m}$\\
$\Varid{allTP}\;\Varid{s}\mathrel{=}\Varid{hfoldr}\;(\Conid{A}\;\Varid{f}\;\Varid{z})$\\
$\hskip0.50em\relax\mathbf{where}$\\
$\hskip0.50em\relax\hskip0.50em\relax\Varid{f}\;\Varid{h}\;\Varid{t}\mathrel{=}\Conid{S}\;(\mathbf{do}\;\Varid{t'}\leftarrow \Varid{unS}\;\Varid{t}$\\
$\hskip0.50em\relax\hskip0.50em\relax\phantom{\Varid{f}\;\Varid{h}\;\Varid{t}\mathrel{=}\Conid{S}\;(\mathbf{do}\;\mbox{}}\Varid{h'}\leftarrow \Varid{applyTP}\;\Varid{s}\;\Varid{h}$\\
$\hskip0.50em\relax\hskip0.50em\relax\phantom{\Varid{f}\;\Varid{h}\;\Varid{t}\mathrel{=}\Conid{S}\;(\mathbf{do}\;\mbox{}}\Varid{return}\;(\Conid{I}\;((\Varid{unI}\;\Varid{t'})\;\Varid{h'})))$\\
$\hskip0.50em\relax\hskip0.50em\relax\Varid{z}\mathrel{=}\Conid{S}\mathbin{\circ}\Varid{return}\mathbin{\circ}\Conid{I}$
\end{tabbing}\snipheader{Try to process the children until one succeeds; preserve the outermost constructor}\begin{tabbing}
\qquad\=\hspace{\lwidth}\=\hspace{\cwidth}\=\+\kill
$\Varid{oneTP}\mathbin{::}\Conid{MonadPlus}\;\Varid{m}\Rightarrow \Conid{TP}\;\Varid{m}\to \Conid{TP}\;\Varid{m}$\\
$\Varid{oneTP}\;\Varid{s}\mathrel{=}\Conid{G}\;(\lambda \Varid{x}\to \Conid{S}\;(\Varid{apply}\;(\Varid{poneTP}\;\Varid{s})\;\Varid{x}\bind $\\
$\phantom{\Varid{oneTP}\;\Varid{s}\mathrel{=}\Conid{G}\;(\lambda \Varid{x}\to \Conid{S}\;(\mbox{}}\Varid{maybe}\;\Varid{mzero}\;(\Varid{return}\mathbin{\circ}\Conid{I})\mathbin{\circ}\Varid{fst}\mathbin{\circ}\Varid{unD}))$
\end{tabbing}
\snipheader{The Duplicate type constructor for encoding type-preserving paramorphisms}\begin{tabbing}
\qquad\=\hspace{\lwidth}\=\hspace{\cwidth}\=\+\kill
$\mathbf{newtype}\;\Conid{D}\;\Varid{t}\;\alpha\mathrel{=}\Conid{D}\;(\Varid{t}\;\alpha,\Varid{t}\;\alpha)$\\
$\Varid{unD}\;(\Conid{D}\;\Varid{x})\mathrel{=}\Varid{x}$
\end{tabbing}
\snipheader{Paramorphic helper for \emph{oneTP}}\begin{tabbing}
\qquad\=\hspace{\lwidth}\=\hspace{\cwidth}\=\+\kill
$\Varid{poneTP}\mathbin{::}\Conid{MonadPlus}\;\Varid{m}\Rightarrow \Conid{TP}\;\Varid{m}\to \Conid{MG}\;(\Conid{D}\;\Conid{Maybe})\;\Varid{m}$\\
$\Varid{poneTP}\;\Varid{s}\mathrel{=}\Varid{hfoldr}\;(\Conid{A}\;\Varid{f}\;\Varid{z})$\\
$\hskip0.50em\relax\mathbf{where}$\\
$\hskip0.50em\relax\hskip0.50em\relax\Varid{f}\;\Varid{h}\;\Varid{t}\mathrel{=}\Conid{S}\;(\mathbf{do}\;\{\mskip1.5mu \Varid{t'}\leftarrow \Varid{unS}\;\Varid{t};$\\
$\hskip0.50em\relax\hskip0.50em\relax\phantom{\Varid{f}\;\Varid{h}\;\Varid{t}\mathrel{=}\Conid{S}\;(\mathbf{do}\;\{\mskip1.5mu \mbox{}}\Varid{tp}\leftarrow \Varid{maybe}\;\Varid{mzero}\;\Varid{return}\;(\Varid{snd}\;(\Varid{unD}\;\Varid{t'}));$\\
$\hskip0.50em\relax\hskip0.50em\relax\phantom{\Varid{f}\;\Varid{h}\;\Varid{t}\mathrel{=}\Conid{S}\;(\mathbf{do}\;\{\mskip1.5mu \mbox{}}(\mathbf{do}\;\Varid{tc}\leftarrow \Varid{maybe}\;\Varid{mzero}\;\Varid{return}\;(\Varid{fst}\;(\Varid{unD}\;\Varid{t'}))$\\
$\hskip0.50em\relax\hskip0.50em\relax\phantom{\Varid{f}\;\Varid{h}\;\Varid{t}\mathrel{=}\Conid{S}\;(\mathbf{do}\;\{\mskip1.5mu \mbox{}}\phantom{(\mathbf{do}\;\mbox{}}\Varid{return}\;(\Conid{D}\;(\Conid{Just}\;(\Varid{tc}\;\Varid{h}),\Conid{Just}\;(\Varid{tp}\;\Varid{h})))$\\
$\hskip0.50em\relax\hskip0.50em\relax\phantom{\Varid{f}\;\Varid{h}\;\Varid{t}\mathrel{=}\Conid{S}\;(\mathbf{do}\;\{\mskip1.5mu \mbox{}})\mathbin{`\Varid{mplus}`}($\\
$\hskip0.50em\relax\hskip0.50em\relax\phantom{\Varid{f}\;\Varid{h}\;\Varid{t}\mathrel{=}\Conid{S}\;(\mathbf{do}\;\{\mskip1.5mu \mbox{}}\phantom{)\mbox{}}\mathbf{do}\;\Varid{h'}\leftarrow \Varid{applyTP}\;\Varid{s}\;\Varid{h}$\\
$\hskip0.50em\relax\hskip0.50em\relax\phantom{\Varid{f}\;\Varid{h}\;\Varid{t}\mathrel{=}\Conid{S}\;(\mathbf{do}\;\{\mskip1.5mu \mbox{}}\phantom{)\mbox{}}\phantom{\mathbf{do}\;\mbox{}}\Varid{return}\;(\Conid{D}\;(\Conid{Just}\;(\Varid{tp}\;\Varid{h'}),\Conid{Just}\;(\Varid{tp}\;\Varid{h})))$\\
$\hskip0.50em\relax\hskip0.50em\relax\phantom{\Varid{f}\;\Varid{h}\;\Varid{t}\mathrel{=}\Conid{S}\;(\mathbf{do}\;\{\mskip1.5mu \mbox{}})\mathbin{`\Varid{mplus}`}\Varid{return}\;(\Conid{D}\;(\Conid{Nothing},\Conid{Just}\;(\Varid{tp}\;\Varid{h})))\mskip1.5mu\})$\\
$\hskip0.50em\relax\hskip0.50em\relax\Varid{z}\;\Varid{x}\mathrel{=}\Conid{S}\;(\Varid{return}\;(\Conid{D}\;(\Conid{Nothing},\Conid{Just}\;\Varid{x})))$
\end{tabbing}\snipheader{Process all children; reduce the intermediate results}\begin{tabbing}
\qquad\=\hspace{\lwidth}\=\hspace{\cwidth}\=\+\kill
$\Varid{allTU}\mathbin{::}\Conid{Monad}\;\Varid{m}\Rightarrow (\Varid{u}\to \Varid{u}\to \Varid{u})\to \Varid{u}\to \Conid{TU}\;\Varid{u}\;\Varid{m}\to \Conid{TU}\;\Varid{u}\;\Varid{m}$\\
$\Varid{allTU}\;\Varid{op2}\;\Varid{unit}\;\Varid{s}\mathrel{=}\Varid{hfoldr}\;(\Conid{A}\;\Varid{f}\;\Varid{z})$\\
$\hskip0.50em\relax\mathbf{where}$\\
$\hskip0.50em\relax\hskip0.50em\relax\Varid{f}\;\Varid{h}\;\Varid{t}\mathrel{=}\Conid{S}\;(\mathbf{do}\;a\leftarrow \Varid{unS}\;\Varid{t}$\\
$\hskip0.50em\relax\hskip0.50em\relax\phantom{\Varid{f}\;\Varid{h}\;\Varid{t}\mathrel{=}\Conid{S}\;(\mathbf{do}\;\mbox{}}\Varid{b}\leftarrow \Varid{applyTU}\;\Varid{s}\;\Varid{h}$\\
$\hskip0.50em\relax\hskip0.50em\relax\phantom{\Varid{f}\;\Varid{h}\;\Varid{t}\mathrel{=}\Conid{S}\;(\mathbf{do}\;\mbox{}}\Varid{return}\;(\Conid{C}\;(\Varid{unC}\;a\mathbin{`\Varid{op2}`}\Varid{b})))$\\
$\hskip0.50em\relax\hskip0.50em\relax\Varid{z}\mathrel{=}\Conid{S}\mathbin{\circ}\Varid{return}\mathbin{\circ}\Conid{C}\mathbin{\circ}\Varid{const}\;\Varid{unit}$
\end{tabbing}\snipheader{Try to process the children until one succeeds; return the processed child}\begin{tabbing}
\qquad\=\hspace{\lwidth}\=\hspace{\cwidth}\=\+\kill
$\Varid{oneTU}\mathbin{::}\Conid{MonadPlus}\;\Varid{m}\Rightarrow \Conid{TU}\;\Varid{u}\;\Varid{m}\to \Conid{TU}\;\Varid{u}\;\Varid{m}$\\
$\Varid{oneTU}\;\Varid{s}\mathrel{=}\Varid{hfoldr}\;(\Conid{A}\;\Varid{f}\;\Varid{z})$\\
$\hskip0.50em\relax\mathbf{where}$\\
$\hskip0.50em\relax\hskip0.50em\relax\Varid{f}\;\Varid{h}\;\Varid{t}\mathrel{=}\Conid{S}\;(\Varid{cast}\;(\Varid{unS}\;\Varid{t})\mathbin{`\Varid{mplus}`}\Varid{cast}\;(\Varid{apply}\;\Varid{s}\;\Varid{h}))$\\
$\hskip0.50em\relax\hskip0.50em\relax\Varid{z}\mathrel{=}\Conid{S}\mathbin{\circ}\Varid{const}\;\Varid{mzero}$\\
$\hskip0.50em\relax\hskip0.50em\relax\Varid{cast}\;\Varid{c}\mathrel{=}\Varid{c}\bind \Varid{return}\mathbin{\circ}\Conid{C}\mathbin{\circ}\Varid{unC}$
\end{tabbing}%
}{%
Examples of one-layer traversal combinators%
}{%
F:allone%
}{}

\myparagraph{One-layer traversal combinators}
Based on \emph{hfoldr}, one can define all kinds of one-layer traversal
combinators. We indicate an open-ended list of candidates by
reconstructing combinators that were presented
elsewhere~\cite{VBT98,LV00,LV02-PADL,Laemmel02-TGT}:
\begin{description}
\item[\emph{allTP}] Process all children; preserve the constructor.
\item[\emph{oneTP}] Try to process the children until one succeeds;
preserve the constructor.
\item[\emph{allTU}] Process all children; reduce the intermediate
results.
\item[\emph{oneTU}] Try to process the children until one succeeds;
return the processed child.
\end{description}
One can think of further combinators, e.g., combinators dealing with
different orders of processing children, monadic effects, and
different constraints regarding success and failure behaviour.

\medskip

\noindent
The definition of the combinators is shown in Fig.~\ref{F:allone}.
Let us explain the first one, that is, \emph{allTP}. We construct a
(type-preserving) fold algebra $A\ f\ z$ and pass it to
\emph{hfoldr}. The ingredient $z$ for the empty constructor
application simply \emph{return}s, i.e., preserves the constructor.
Additional wrapping of $S$ and $I$ is needed to deal with our
`datatypes-as-type-constructors' encoding. The ingredient $f$ sequences
three computations. Firstly, the tail $t$ is computed. Secondly, the
head $h$ (i.e., the child at hand) is processed via $s$. Thirdly, the
processed head $h'$ is passed to the computed tail $t'$ (modulo
wrapping and unwrapping). The definition of the other combinators is
similar. A challenging complication shows up when we want to define
\emph{oneTP} because it turns out that this fold over children is
paramorphic in nature~\cite{Meertens92} while our combinator
\emph{hfoldr} is catamorphic. We use a pairing technique adopted
from~\cite{Meertens92} to encode the paramorphic fold with
\emph{hfoldr} (cf.\ the new type constructor $D$). This complication
also illustrates the potential for further generic function types,
namely $\mathit{MG}\ (D\ \mathit{Maybe})\ m$ in this case.

\newpage

\myparagraph{Sums of products vs.\ curried constructor applications}
The combinator \emph{hfoldr} roughly covers the sum and product cases
in a polytypic definition~\cite{JJ96} but without resorting to a
constructor-free representation type. While a polytypic definition
deals with the many children in a term via nested binary products,
\emph{hfoldr} processes all children in a curried constructor
application as is. While the different constructors of a datatype
amount to nested binary sums in a polytypic definition, the base case
of \emph{hfoldr} handles the empty constructor application. Also note
that polytypic definitions employ a designated declaration form for
type induction while functional strategies are defined as ordinary
recursive functions. Finally, polytypic definitions are normally
implemented by compile-time specialisation. Functional strategies only
require a \emph{Term} interface for datatypes to enable \emph{hfoldr}
and \emph{adhoc}.

%%%%%%%%%%%%%%%%%%%%%%%%%%%%%%%%%%%%%%%%%%%%%%%%%%%%%%%%%%%%%%%%%%%%%%%%%%%%%%

\section{Largo}

We are now in the position to define all kinds of traversal schemes in
terms of the strategy combinators and recursion. Before we present a
portfolio of such schemes, we impose some extra structure on our
combinator suite in order to allow for a uniform treatment of
type-preserving and type-unifying traversal.

\myparagraph{Overloading \emph{TP} and \emph{TU}}
The strategy combinators are usually typed in a manner that they can
be used to derive strategies of both type \emph{TP} and type
\emph{TU}. Hence, one would expect that traversal schemes can be
composed in a manner to work for both type schemes as well. There are
two asymmetries which we need to address. Firstly, there is no obvious
way to write fold algebras that cover both \emph{TP} and
\emph{TU}. This is the reason why we have a \emph{oneTP} vs.\ a
\emph{oneTU}, and an \emph{allTP} vs.\ an \emph{allTU}. These
combinators should be overloaded. Secondly, there is no uniform way to
\emph{combine} the execution of two strategies of the \emph{same}
type. We can only compose \emph{alternative} branches via
\emph{choice}. For the sake of sequential composition being generic,
the \emph{seq} combinator insists on a type-preserving first
argument. For the sake of value passing to deliver an extra input to
the second operand, the \emph{pass} combinator insists on a
type-unifying first argument. We suggest an overloaded composition
combinator \emph{comb} as follows. In the \emph{TP} instance,
sequential composition via \emph{seq} is appropriate: the term as
returned by the first type-preserving application is passed to the
second. In the \emph{TU} instance, both strategies are evaluated via
\emph{pass}, and then, the two resulting values are combined via the
binary operation of a monoid. Note that the \emph{Monoid} class is
also appropriate to harmonise the types of the one-layer traversal
combinators \emph{allTP} and \emph{allTU} for overloading. That is,
the additional parameters for \emph{allTU} are assumed to correspond
to the monoid operations.

\medskip

\noindent
In Fig.~\ref{F:overloading}, the overloaded combinators for
traversal (cf.\ \emph{one} and \emph{all}) and combination of
strategies (cf.\ \emph{comb}) are defined. For completeness, we also
overload \emph{idTP} and \emph{constTU} in a way that an always
succeeding strategy is defined (cf.\ \emph{skip}). The overloaded
combinators are placed in two Haskell type classes
\emph{StrategicMonadPlus} and \emph{StrategicMonoid} in order to deal with
the different class constraints for the overloaded combinators.

\myfigure{
\input{snip/overloading.math}%
}{%
Classes of generic function combinators
}{%
F:overloading%
}{}

\myparagraph{Full traversal control}
In Fig.~\ref{F:schemes}, we define traversal schemes in three
groups. The highly parameterised function \emph{traverse} captures a
rich class of traversal schemes as illustrated by the given
instantiations. Some of the type-preserving instantiations of
\emph{traverse} were identified in \cite{VBT98}. In
\cite{LV00,Laemmel02-TGT,LV02-PADL}, some type-unifying instantiations
were added. In the definition of \emph{traverse}, we separate out the
distinguishing parameters of such more specific traversal schemes. The
argument \emph{op} is the function combinator to compose term
processing and recursive descent. The argument $t$ defines how to
descend into children. Finally, $f$ is the strategy for term
processing. Here is the normative type for this `mother of traversal':

\input{snip/mother.math}

\myfigure{
\input{snip/schemes.math}
}{%
Traversal schemes%
}{%
F:schemes%
}{}

\noindent
We call this a \emph{normative} type because the inferable type is
much more general, namely $\mathit{traverse} :: (a \to b \to c) \to (c
\to b) \to a \to c$. Not even the normative type captures all our
intentions, e.g., the one that $t$ actually specifies descent into
children. It is a topic for future work to capture such constraints in
suitable types.

\medskip

\noindent
The scheme \emph{traverse} is illustrated by several instances in the
figure. The first two instances \emph{all\_rec} and \emph{one\_rec}
refine \emph{traverse} to opt either for descent with \emph{all} or
\emph{one}, respectively.  The postfixes ``...\emph{td}'' or
``...\emph{bu}'' stand for top-down or bottom-up, respectively.  The
prefixes ``\emph{full}...'', ``\emph{once}...'', or ``\emph{stop}...''
hint on the style of traversal control: full traversal where all nodes
are processed vs.\ single hit traversal where the argument strategy
has to succeed once vs.\ cut-off traversal where descent stops when a
subterm was processed successfully. All the more concrete schemes are
still overloaded as for the choice of \emph{TP} vs.\ \emph{TU}. To
give an example, the scheme \emph{full\_td} models full traversal of a
term where all nodes are processed in top-down manner. In this
example, the argument $t$ for descent is instantiated with \emph{all}
(via \emph{all\_rec}). The argument \emph{op} for the composition of
term processing and recursive descent is instantiated with
\emph{comb}.

\myparagraph{Traversal with propagation}
The second group in Fig.~\ref{F:schemes} illustrates a class of
traversal schemes that go beyond simple node-wise processing. In
addition to traversal, \emph{p}ropagation of an \emph{e}nvironment
(abbreviated by the postfix ``...\emph{pe}'') is performed.  The
scheme \emph{propagate} takes the same arguments \emph{op}, $t$, and
$f$ as \emph{traverse}.  In addition, the scheme carries a generic
function argument $u$ to update the environment before descent into
the children, and an argument $e$ for the initial environment. Here is
the normative type for \emph{propagate}:

\newpage

\input{snip/pe.math}

\noindent
The shown instances of \emph{propagate} favour top-down traversal
because bottom-up traversal seems to be less obvious when combined
with propagation. In is interesting to notice that the scheme
\emph{propagate} provides an alternative to using the environment
monad~\cite{Wadler92} in a strategic program.

\myparagraph{Path schemes}
The last group in Fig.~\ref{F:schemes} deals with traversal
constrained by the path leading to or starting from a node (say, a
subterm). For this reason, we do not just use a processor strategy
$f$, but we also use predicates $p$ for constraints. The scheme
\emph{beloweq} attempts to process a subterm via $f$ (not necessarily
strictly) below a node for which $p$ succeeds; dually for
\emph{aboveeq}. The \emph{below} and \emph{above} variants enforce the
root of the processed subterm not to coincide with the constrained
node. In the definition of the path schemes, we use the simpler
traversal schemes \emph{once\_td} and \emph{once\_bu} from above. The
types of the path schemes are as follows:

\input{snip/path.math}

\noindent
These types illustrate that predicates are encoded as type-unifying
functions of type $\mathit{TU}\ ()\ m$ where $m$ is normally the
\emph{Maybe} monad or another instance of the \emph{MonadPlus}
class. So success and failure behaviour encodes the truth value.

\medskip

\noindent
The final three schemes in the group elaborate the more basic path
schemes to deal with \emph{lists} of predicates. The schemes
\emph{belowlist} and \emph{abovelist} constrain the prefix or postfix
of a node rooting a subtree of interest, respectively. The last scheme
\emph{prepost} combines \emph{belowlist} and \emph{abovelist}. Note
that the path scheme \emph{belowlist} resolves the remaining blind
spot in the sample code for our running example in
Fig.~\ref{F:running}.

%%%%%%%%%%%%%%%%%%%%%%%%%%%%%%%%%%%%%%%%%%%%%%%%%%%%%%%%%%%%%%%%%%%%%%%%%%%%%%

\medskip

\hfill{\textsf{[End Of Symphony]}}

%%%%%%%%%%%%%%%%%%%%%%%%%%%%%%%%%%%%%%%%%%%%%%%%%%%%%%%%%%%%%%%%%%%%%%%%%%%%%%

\newpage

\mysection{Appraisal}

\medskip

\noindent
The four movements of this symphony combine certain bits of parametric
polymorphism, type case (i.e., intensional polymorphism), polytypism,
and overloading to provide a concise and expressive model for
functional strategies~\cite{LV02-PADL}. The development culminates in
the last movement when typed highly parameterised traversal schemes
are defined. This paper makes the following overall contributions when
compared to previous work on strategic programming. Firstly, the
developed model is more concise and expressive than in our previous
work~\cite{LV00,Laemmel02-TGT,LV02-PADL}. Secondly, the defined
traversal schemes reach new limits of parameterisation. We shall
elaborate on these two claims accordingly.

\medskip

\noindent
The overall model ``functional strategies = first-class polymorphic
functions'' was already sketched in~\cite{LV02-PADL}. In addition to
working out this model, we improved it in two important ways. Firstly,
the type schemes for type-preserving and type-unifying strategies are
viewed as instances of one common type scheme \emph{Strategic}.  This
allows us to define combinators like \emph{choice} and \emph{adhoc}
without commitment to any specific type scheme. Secondly, we
eliminated the need for an open-ended list of primitive one-layer
traversal combinators. To this end, we designed a generic fold
combinator for processing the immediate subterms of constructor
applications.  A related language construct is described
in~\cite{DV01}: pattern matching is generalised in a way that generic
access is granted to subterms of constructor applications. A limited
form of folds over constructor applications was first proposed in
\cite{Laemmel02-TGT} in a term-rewriting setting, but there it was by
far less potent and it was difficult to type due to the limitations of
the underlying type system.

\medskip

\noindent
The key idea to define traversal schemes in terms of one-layer
traversal combinators carries over from the seminal work on term
rewriting strategies~\cite{VBT98}. The present paper reaches a new
level of genericity for the following two reasons. Firstly, we
effectively overload the two prime type schemes of generic
traversal. This allows us to capture meaningful traversal schemes
without anticipating their use for transformation vs.\ analysis.
Secondly, we identify a few highly parameterised traversal schemes
with parameters for various forms of control. These `skeleton' or
`meta' schemes capture more concrete traversal schemes favoured in
previous work.

\myparagraph{Acknowledgement}
The following people contributed beats to this polymorphic symphony:
Jan Kort, Andres L{\"o}h, Simon Peyton-Jones, Joost Visser, and
Stephanie Weirich. I am grateful for the remarks and suggestions by
the anonymous WRS'02 referees. I devote my first symphony to Berenike
and Bernadette.

%%%%%%%%%%%%%%%%%%%%%%%%%%%%%%%%%%%%%%%%%%%%%%%%%%%%%%%%%%%%%%%%%%%%%%%%%%%%%%

\bibliographystyle{abbrv}
\bibliography{paper}

%%%%%%%%%%%%%%%%%%%%%%%%%%%%%%%%%%%%%%%%%%%%%%%%%%%%%%%%%%%%%%%%%%%%%%%%%%%%%%

\end{document}